\documentclass[pra,twocolumn,preprintnumbers,superscriptaddress]{revtex4}
\usepackage{amssymb}
\usepackage{amsmath}
\usepackage{amsfonts}
\usepackage{graphicx}
\usepackage{dcolumn}
\usepackage{bm}
\usepackage{verbatim}
\usepackage{color}
\usepackage{wasysym}

\begin{document}

\title{Effective potential and superfluidity of microwave-dressed polar molecules}
\date{\today }
\author{Fulin Deng}
\affiliation{School of Physics and Technology, Wuhan University, Wuhan, Hubei 430072, China}
\affiliation{CAS Key Laboratory of Theoretical Physics, Institute of Theoretical Physics, Chinese Academy of Sciences, Beijing 100190, China}

\author{Xing-Yan Chen}
\affiliation{Max-Planck-Institut f\"ur Quantenoptik, 85748 Garching, Germany}
\affiliation{Munich Center for Quantum Science and Technology, 80799 M\"unchen, Germany}

\author{Xin-Yu Luo}
\affiliation{Max-Planck-Institut f\"ur Quantenoptik, 85748 Garching, Germany}
\affiliation{Munich Center for Quantum Science and Technology, 80799 M\"unchen, Germany}

\author{Wenxian Zhang}
\affiliation{School of Physics and Technology, Wuhan University, Wuhan, Hubei 430072, China}
\affiliation{Wuhan Institute of Quantum Technology, Wuhan, Hubei 430206, China}

\author{Su Yi}
\email{syi@itp.ac.cn}
\affiliation{CAS Key Laboratory of Theoretical Physics, Institute of Theoretical Physics, Chinese Academy of Sciences, Beijing 100190, China}
\affiliation{CAS Center for Excellence in Topological Quantum Computation \& School of Physical Sciences, University of Chinese Academy of Sciences, Beijing 100049, China}
\affiliation{Peng Huanwu Collaborative Center for Research and Education, Beihang University, Beijing 100191, China}

\author{Tao Shi}
\email{tshi@itp.ac.cn}
\affiliation{CAS Key Laboratory of Theoretical Physics, Institute of Theoretical Physics, Chinese Academy of Sciences, Beijing 100190, China}
\affiliation{CAS Center for Excellence in Topological Quantum Computation \& School of Physical Sciences, University of Chinese Academy of Sciences, Beijing 100049, China}
\affiliation{Peng Huanwu Collaborative Center for Research and Education, Beihang University, Beijing 100191, China}

\begin{abstract}
For microwave-dressed polar molecules, we analytically derive an intermolecular potential composed of an anisotropic van der Waals shielding core and a long-range dipolar interaction. We validate this effective potential by comparing its scattering properties with those calculated using the full multi-channel interaction potential. It is shown that scattering resonances can be induced by a sufficiently strong microwave field. We also show the power of the effective potential in the study of many-body physics by calculating the critical temperature of the Bardeen-Cooper-Schrieffer pairing in the microwave-dressed NaK gas. It turns out that the effective potential is well-behaved and extremely suitable for studying the many-body physics of the molecular gases. Our results pave the way for the studies of the many-body physics of the ultracold microwave-dressed molecular gases.
\end{abstract}

\maketitle

\textit{Introduction.}---Ultracold gases of polar molecules~\cite{Carr2009,Ye2017} provide a unique platform for the exploration of quantum information~\cite{Zoller2006a}, quantum computing \cite{DeMille2002,Cornish2020}, quantum simulation~\cite{Zoller2006,Zwierlein2021}, quantum chemistry~\cite{Krem2008,Ni2019}, and precision measurement~\cite{Kozlov2007,Berger2010,Hinds2011}. From the condensed-matter perspective, the strong long-range and anisotropic dipole-dipole interaction (DDI) make ultracold polar molecules an ideal platform for investigating strongly correlated many-body physics~\cite{Pfau2009,Baranov2012}. Over the past decade, there are tremendous experimental efforts for the creation of the ultracold molecular gases by both direct cooling~\cite{Tarbutt2021} and cold-atom assembly. Particularly, indirect production of the high-phase-space-density molecular gases from ultracold atomic gases via the Feshbach resonance and stimulated Raman adiabatic passage has been successfully employed to create bialkali molecules of KRb~\cite{Ye2008}, RbCs \cite{Nagerl2014,Cornish2014}, NaK \cite{Zwierlein2015,Bloch2018,Ospelkaus2020}, NaRb \cite{Wang2016}, NaLi \cite{Jamison2017}, and NaCs~\cite{Will2022,Ni2021}. Recently, starting from the association of double degenerate Bose-Fermi mixtures~\cite{Ye2019,Luo2021} and evaporative cooling~\cite{Ye2020a,Ye2021,Luo2022a} enabled by the collisional shielding with either a d.c.~\cite{Ye2020a,Ye2021,Ye2020b} or a microwave field~\cite{Luo2022a,Doyle2021}, degenerate Fermi gases of polar molecules have finally become available in experiments.

Unlike the conventional DDI induced by a d.c. electric field, the long-range DDI between microwave-shielded molecules in the highest dressed state is attractive in the plane of the microwave field~\cite{Luo2022a}, which may lead to exotic $p$-wave superfluids~\cite{You1999,Baranov2002,Shlyapnikov2009,Shi2010,Shlyapnikov2011}. Because DDI couples different rotational states, a complete description of the intermolecular interaction involves multiple dressed rotational states of the molecules~\cite{Karman2018,Karman2022}, which is cumbersome for the studies of the many-body physics of a single shielded dressed state. Therefore, a simple and accurate effective potential is an essential ingredient for exploring the many-body physics of molecular gases.

In this Letter, we analytically derive an effective potential between two microwave-dressed polar molecules. At large inter-molecular distance, this potential is a negated DDI such that it is attractive in the plane of the microwave field and repulsive along the propagation direction of microwave. While at short range, the potential is of the $1/r^6$ type and is anisotropically repulsive. As a result, the effective potential has a shielding core in all three dimensions. The validity of this effective potential is justified by comparing it with numerically obtained adiabatic potential and by exploring the scattering properties of two molecules. We show that the effective potential not only leads to the correct scattering cross sections, but also accurately predicts the position of the scattering resonance. Finally, as an application of the effective potential, we study the Bardeen-Cooper-Schrieffer (BCS) superfluidity in the microwave-dressed NaK gas, where the Rabi-frequency of the microwave field plays the role as a control knob to tune the superfluid critical temperature. It turns out that the effective potential is well-behaved and suitable for studying the many-body physics of molecular gases.

\begin{figure}[tbp]
\includegraphics[width=1\linewidth]{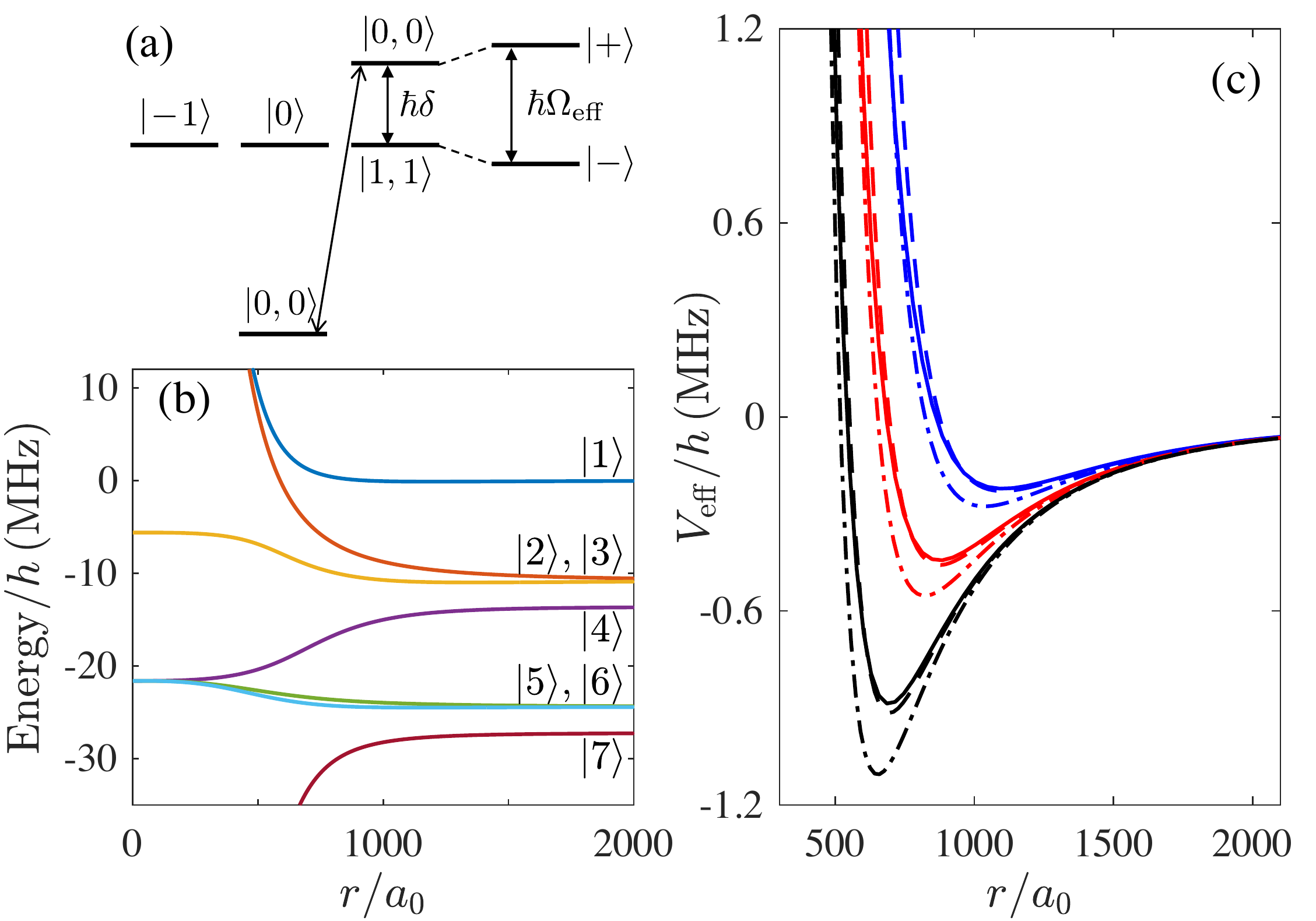}
\caption{(a) Schematic of the level structure of a microwave-dressed molecule. (b) Typical adiabatic potential curves of two colliding molecules for seven dressed state channels. (c) Effective potentials along $\theta=\pi/2$ obtained by numerical diagonalization (solid lines), numerical fitting (dashed line), and analytical expressions (dash-dotted lines) for $\delta_r=0.1$ and $\Omega/(2\pi)=20$, $50$, and $80\,{\rm MHz}$ (for three sets of curves in descending order).}
\label{figC6}
\end{figure}

\textit{Effective molecule-molecule interaction}.---We consider a gas of the NaK molecules in the ${}^1\Sigma(v=0)$ state which exhibits a molecular-frame dipole moment $d= 2.72$ Debye. Under ultracold temperature, only the rotational degree of freedom is relevant such that the Hamiltonian of a single molecule is $\hat h_{\rm rot}=B_{\rm rot}{\mathbf J}^2$, where $B_{\rm rot}/\hbar=2\pi\times 2.822\,\mathrm{GHz}$ is the rotational constant and ${\mathbf J}$ is the angular momentum operator. Since the rotation spectrum, $B_{\rm rot}J(J+1)$, is anharmonic, we focus on the two lowest rotational manifolds ($J=0$ and $1$) which are split by an energy $\hbar\omega_e=2B_{\rm rot}$. Correspondingly, the Hilbert space for the internal states of a molecule is defined by four states: $|J,M_J\rangle=|0,0\rangle$, $|1,0\rangle$, and $|1,\pm1\rangle$. To achieve the microwave shielding, molecules with electric dipole moment $d_0\hat{\mathbf d}$ are illuminated by a position-independent $\sigma^{+}$-polarized microwave propagating along the z axis, where $\hat{\mathbf d}$ is the unit vector along the internuclear axis of the molecule and the microwave field rotates circularly in the $xy$ plane with frequency $\omega_0$. Within the internal-state Hilbert space, the coupling between the microwave and the molecular rotational states gives rise to the Hamiltonian $\hat h_{\rm mw}=\frac{\hbar\Omega}{2}e^{-i\omega_0t}|1,1\rangle\langle 0,0|+{\rm h.c.}$, where $\Omega$ is the Rabi frequency. Then, in the interaction picture, the eigenstates of the internal-state Hamiltonian, $\hat h_{\rm in}=\hat h_{\rm rot}+\hat h_{\rm mw}$, are $|0\rangle\equiv|1,0\rangle$, $|-1\rangle\equiv |1,-1\rangle$, $|+\rangle\equiv u|0,0\rangle + v|1,1\rangle$, and $|-\rangle\equiv u|1,1\rangle- v|0,0\rangle$, where $u=\sqrt{(1-\delta /\Omega_{\mathrm{eff}})/2}$ and $v=\sqrt{(1+\delta /\Omega _{\mathrm{eff}})/2}$ with $\delta =\omega _{e}-\omega _{0}$ being the detuning and $\Omega_{\mathrm{eff}}=\sqrt{\delta ^{2}+\Omega ^{2}}$ the effective Rabi frequency. The corresponding eigenenergies are $E_{0}=E_{-1}=\delta$ and $E_{\pm}=(\delta \pm \Omega _{\mathrm{eff}})/2$. Figure~\ref{figC6}(a) schematically shows the level structure of a molecule.

For two molecules with dipole moments $d\hat{\mathbf d}_1$ and $d\hat{\mathbf d}_2$, the dipole-dipole interaction (DDI) between them is
\begin{align}
V({\mathbf r})=\frac{d^2}{4\pi\epsilon_{0}r^3}\left[\hat{\mathbf d}_1\cdot \hat{\mathbf d}_2-3(\hat{\mathbf d}_1\cdot\hat{\mathbf r})(\hat{\mathbf d}_2\cdot\hat{\mathbf r})\right],
\end{align}
where $\epsilon_{0}$ is the electric permittivity of vacuum, $r=|{\mathbf r}|$, and $\hat{\mathbf r}={\mathbf r}/r$. To express DDI in the two-molecule internal Hilbert space, we note that the two-particle Hamiltonian $\hat H_2=\sum_{j=1,2}\hat h_j+V({\mathbf r}_1-{\mathbf r}_2)$ possesses a parity symmetry, where $\hat h_j=-\hbar^2\nabla_j^2/(2M)+\hat h_{\rm in}(j)$ with $M$ being the mass of the molecule. This suggests that the symmetric and antisymmetric two-particle internal states are decoupled in the Hamiltonian $\hat H_2$. Here we focus on the ten-dimensional symmetric subspace in which the shielding states of the molecules lie. It turns out that, under the rotating-wave approximation, $V({\mathbf r})$ in the seven-dimensional (7D) symmetric subspace, $\mathcal{S}_7\equiv \{|\nu\rangle\}_{\nu=1}^7$, is decoupled from the remaining three-dimensional symmetric subspace, where $|1\rangle=|+,+\rangle$, $|2\rangle=|+,0\rangle_s$, $|3\rangle=|+,-1\rangle_s$, $|4\rangle=|+,-\rangle_s$, $|5\rangle=|-,0\rangle_s$, $|6\rangle=|-,-1\rangle_s$, and $|7\rangle=|-,-\rangle$ with $|i,j\rangle_s=(|i, j\rangle+|j,i\rangle)/\sqrt{2}$. Correspondingly, with respect to the asymptotical state $|\nu=1\rangle$, the energies of these states are $\mathcal{E}_{\nu}=\{0,\frac{1}{2}(\delta-\Omega_{\rm eff}),\frac{1}{2}(\delta-\Omega_{\rm eff}),-\Omega_{\rm eff},\frac{1}{2}(\delta-3\Omega_{\rm eff}),\frac{1}{2}(\delta-3\Omega_{\rm eff}),-2\Omega_{\rm eff}\}$. In below, we shall consider the two-molecule problem only in the subspace $\mathcal{S}_7$.

To derive an effective potential between two molecules, we make use of the Born-Oppenheimer approximation which holds when the kinetic energy of the molecules is much smaller than the energy level spacings between internal states ($\sim \Omega _{\mathrm{eff}}$). After diagonalizing $V(\mathbf{r})$ in $\mathcal{S}_7$, we find seven adiabatic potentials corresponding to different dressed-state channels [see, e.g., Fig.~\ref{figC6}(b) for the typical adiabatic potential curves]. Particularly, the effective potential for two molecules in the dressed state $|+\rangle$ is the highest adiabatic curve. Remarkably, as shown in the Supplemental Material (SM), there exists an approximate expression for the effective potential, i.e.,
\begin{align}
V_{\mathrm{eff}}(\mathbf{r})&=\frac{C_{3}}{r^{3}}P_{2}(\cos\theta)+\frac{C_{6}}{r^{6}}A(\theta),
\label{Vps}
\end{align}%
where $P_l(\cos\theta)$ is the Legendre polynomial with $\theta$ being the polar angle of ${\mathbf r}$ and $A(\theta)=7-5P_{2}(\cos\theta)-2P_{4}(\cos\theta)$. Moreover, $C_{3}=d^{2}/\left[24\pi \epsilon_{0}(1+\delta_r^2)\right]$, $C_6=d^4/\left[1120\pi^2\epsilon_0^2\Omega(1+\delta_r^2)^{3/2}\right]$ with $\delta_r=\delta/\Omega$. The first term of $V_{\mathrm{eff}}$ represents DDI which, different from the conventional one, is attractive in the $xy$ plane and repulsive along the $z$ axis. Because $A(\theta)>0$ when $\theta\neq 0$ or $\pi$, the second term is repulsive and provides a shielding core away from the $z$ axis. Interestingly, even along the $z$ axis on which $A(\theta)$ vanishes, DDI itself is repulsive and prevents two molecules from getting close to each other.

In Fig.~\ref{figC6}(c), the effective potential~\eqref{Vps} is benchmarked by the highest adiabatic curve obtained from diagonalizing $V(\mathbf{r})$. Generally speaking, the expression for $C_3$ is accurate in the sense that it gives rise to the correct long-range behavior, while the analytical expression for $C_6$ is a good approximation only when $\Omega>d_0^2/(4\pi\epsilon_{0}r^3)$. In any case, one can alternatively determine the values of $C_3$ and $C_6$ by fitting the adiabatic potential curve, which, as shown in Fig.~\ref{figC6}(c) and also in SM, yields satisfactory results in the energy range of interest to us. The advantage of Eq.~\eqref{Vps} is that it establishes an intuitive connection between the potential and the physical parameters of the microwave field. In addition, as shall be shown below, this effective potential is well-behaved and can be used for studying the many-body problems.

\textit{Two-body scatterings}.---To further justify the effective potential, we investigate the low-energy scattering of two shielding molecules interacting via $V_{\rm eff}({\mathbf r})$. Since this study only involves a single scattering channel ($\nu=1$ in $\mathcal{S}_7$), its results should be checked by the scattering calculations involving all seven channels. To this end, let us briefly outline the theoretical treatment for the multi-channel scattering~\cite{Karman2018,Karman2022,Bohn2003,Quemener2018}. The Schr\"{o}dinger equations governing the relative motion of two colliding molecules are
\begin{align}
\sum_{\nu^{\prime}=1}^7\left(-\frac{\hbar ^{2}\nabla ^{2}}{M}\delta _{\nu \nu^{\prime}}+V_{\nu \nu^{\prime }}\right)\psi _{\nu ^{\prime }}({\mathbf{r}})=\frac{\hbar^2k_{\nu}^2}{M}\psi_{\nu}({\mathbf{r}}),\label{scase}
\end{align}
where $\psi_{\nu }({\mathbf{r}})$ is the wave function of the $\nu$th scattering channel, $V_{\nu \nu^{\prime }}=\langle \nu |V|\nu ^{\prime }\rangle $, and $k_\nu=\sqrt{k_{1}^{2}-M\mathcal{E}_\nu/\hbar^2}$ is the incident momentum of the $\nu$th scattering channel. To solve Eq.~\eqref{scase}, we first expand the wave functions in the partial-wave basis as $\psi_{\nu}({\mathbf{r}})=\sum_{lm}Y_{lm}(\hat{\mathbf{r}})\phi_{\nu lm}(r)/r$, where $l$ is odd for identical fermions. The equations for $\phi_{\nu lm}$ can be numerically evolved from $r=0$ to a sufficiently large value $r_{\infty}$ using Johnson's log-derivative propagator method~\cite{log-Johnson}. Then, by comparing $\phi_{\nu lm}$ with the asymptotical boundary condition, we obtain the scattering amplitude $f^{\nu'l'm'}_{\nu lm}$ and cross section $\sigma^{\nu'l'm'}_{\nu lm}=4\pi \left|f^{\nu'l'm'}_{\nu lm}\right|^2$ for the $(\nu lm)$ to $(\nu'l'm')$ scattering. It should be noted that $\sigma^{\nu'l'm'}_{\nu lm}$ is nonzero only when $m$ and $m'$ satisfy $m=m_{\nu}$ and $m'=m_{\nu'}$ where, for $\nu=1$ to $7$, $m_{\nu}=m_1,m_1+1,m_1+2,m_1,m_1+1,m_1+2$, and $m_1$, respectively, with $m_1$ being an integer. Numerically, to ensure the convergence of the scattering cross sections, we normally choose $k_{0}r_{\infty}>32$ and $l_c> 11$, where $l_c$ is the truncation imposed on the orbital angular momentum.

As a special case of the multi-channel scattering, the Schr\"odinger equation for single-channel scattering can be obtained by projecting Eq.~\eqref{scase} onto the $\nu=1$ channel with $V_{11}$ being replaced by $V_{\rm eff}$. In addition, we denote the single-channel scattering cross section as $\sigma_{lm}^{l'm'}$.

\begin{figure}[tbp]
\includegraphics[width=0.75\linewidth]{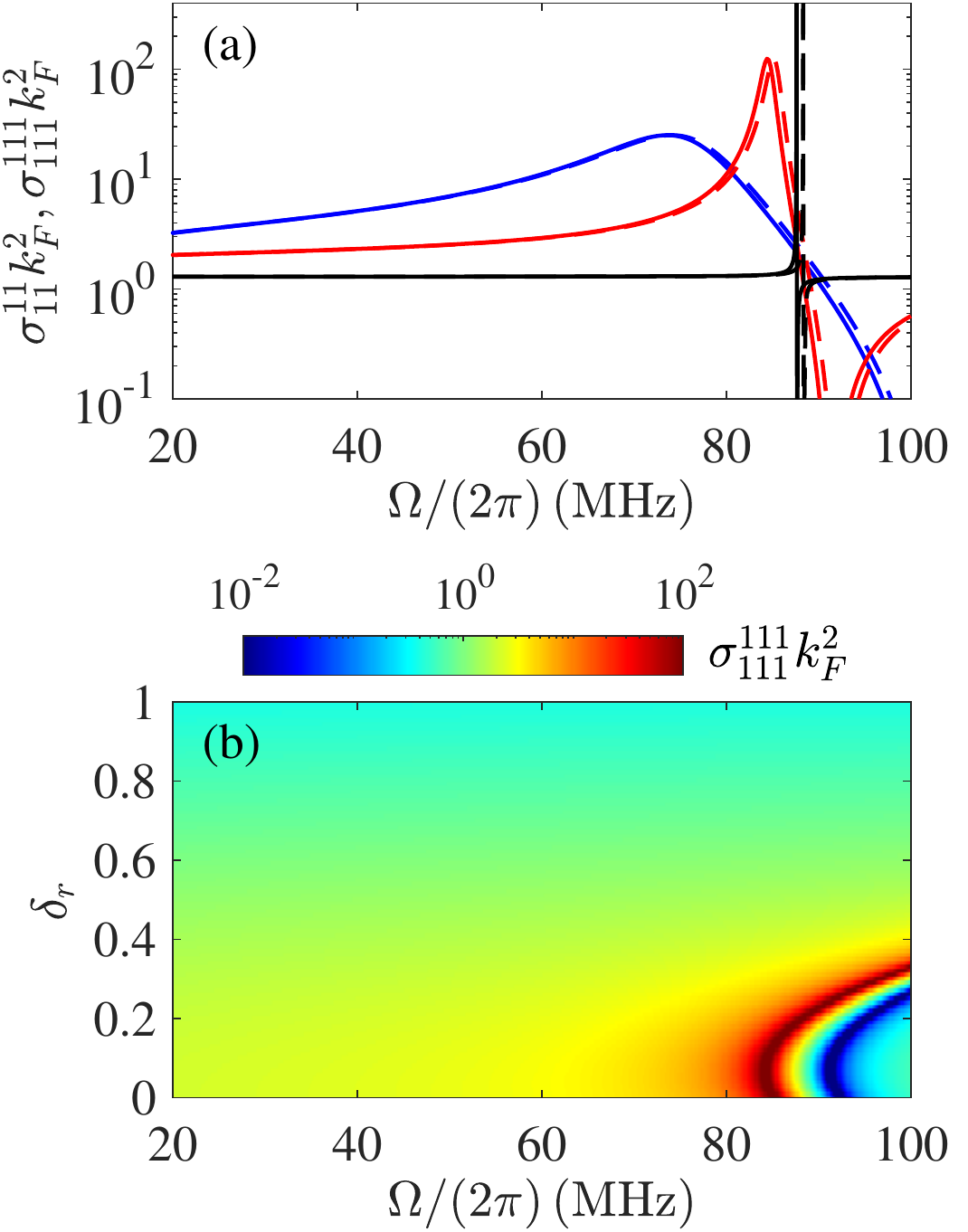}
\caption{(a) $p$-wave scattering cross sections $\sigma_{11}^{11}k_F^2$ (solid lines) and $\sigma_{111}^{111}k_F^2$ (dashed lines) as functions of $\Omega$ for $\delta_r=0.1$ and $k_{1}/k_{F}=0.04$ (black lines), $0.45$ (red lines), and $1$ (blue lines). (b) $\sigma _{111}^{111}k_F^2$ as a function of $\Omega$ and $\delta_r$ for $k_{1}/k_{F}=0.45$.}
\label{figSC}
\end{figure}

Since the scattering cross section of the $p$ wave is dominant over all other partial waves~\cite{SM}, we compare, in Fig.~\ref{figSC}(a), $\sigma_{11}^{11}$ and $\sigma_{111}^{111}$ for $\delta_r=0.1$ and $k_1/k_F=0.04$, $0.45$, and $1$, where $k_F=(6\pi^2n_0)^{1/3}$ is the Fermi wave vector with $n_{0}=10^{12}\,{\rm cm}^{-3}$ being the density of the experimentally realized molecular gas~\cite{Luo2022a}. As can be seen, away from scattering resonances, quantitative agreements have been achieved for $p$-wave cross section under different incident momenta. Moreover, the single-channel calculations can even predict the position of scattering resonance with high accuracy. These results together with other comparisons in SM validate the usage of the effective potential, which, as shown below, significantly simplifies the calculations in the many-body problems.

As to the $\Omega$ dependence of the $p$-wave cross section, it can be seen that, for small $k_1$, $\sigma_{111}^{111}$ barely changes as $\Omega$ varies over a wide range. Then at $\Omega/(2\pi)\approx 87.7\,{\rm MHz}$ a narrow scattering resonance appears, signaling the formation of a quasi-bound state. Furthermore, as $k_1$ increases to $k_F$, the resonance peak shifts to $\Omega/(2\pi)\approx 73.6\,{\rm MHz}$ and the width of the resonance is significantly broadened. To understand these features, let us recall that 1) there is a centrifugal barrier for the $p$-wave potential; 2) a scattering resonance implies that a quasi-bound state with energy in resonance with the incident energy forms inside the barrier. Now, as $k_1$ increases, the resonant quasi-bound state energy also increases and gets closer to the top of the potential barrier. Consequently, the lifetime of the quasi-bound state is shortened due to large decay rate, which leads to a broader resonance. In addition, the increasing quasi-bound state energy implies a weakened attractive interaction via reducing $\Omega$ (see the relation between $C_6$ and $\Omega$). Therefore, the resonance peak shifts towards the lower $\Omega$ direction as $k_1$ increases.

To reveal more details about the scattering resonance, we map out, in Fig.~\ref{figSC}(b), $\sigma_{111}^{111}$ on the $\Omega$-$\delta_r$ parameter plane for $k_1=0.45k_F$. As can be seen, a resonant peak appears in the parameter region $\delta_r\apprle 0.3$ and $\Omega/(2\pi) \apprge 80\,{\rm MHz}$. To give an intuitive explanation to the relation between the resonance and control parameters, a $p$-wave bound state at threshold appears when the WKB phase
\begin{align}
\varphi_p=\int_{v_p(r)\leq0}\sqrt{-Mv_p(r)/\hbar^2} dr\propto \left[\frac{\Omega^2}{(1+\delta_r^2)^5}\right]^{1/12} \label{wkbph}
\end{align}
is sufficiently large, where $v_p(r)=\int d\hat{\mathbf r}|Y_{1m}(\hat{\mathbf r})|^2V_{\rm eff}({\mathbf r})$. Clearly, both increasing $\Omega$ and decreasing $\delta_r$ favor the appearance of a shape resonance and the formation of a bound state.

\textit{Superfluid phase transitions}.---As an application of the effective potential, we now turn to explore the BCS superfluid phase transition in a homogeneous gas of the dressed-state molecules with density $n_0$. In particular, we focus on the transition temperature $T_c$ and the pairing wave functions. Previously, the superfluidity of the fermionic dipolar gases have been extensively studied using the pseudopotential containing a contact part and a bare DDI~\cite{You1999,Baranov2002,Shlyapnikov2009,Baranov2004,Shlyapnikov2011,Marenko2006,Shi2010,Hirsch2010,Pu2010,Shi2014,Zhai2013}. To illustrate the effect of microwave field to the superfluidity, we apply the effective potential to the molecules in the dressed state $|+\rangle$, and write down the many-body Hamiltonian
\begin{align}
\hat H &=\int d^{3}\mathbf{r}\hat{\psi}^{\dagger }(\mathbf{r})\left(-\frac{\hbar^2\nabla ^{2}}{2M}-\mu \right) \hat{\psi}(\mathbf{r}) \nonumber\\
&\quad+\frac{1}{2}\int d\mathbf{r}d\mathbf{r}^{\prime }\hat{\psi}^{\dagger }(\mathbf{r})\hat{\psi}^{\dagger }(\mathbf{r}^{\prime })V_{\mathrm{eff}}(\mathbf{r}-\mathbf{r}^{\prime })\hat{\psi}(\mathbf{r}^{\prime })\hat{\psi}(\mathbf{r}),
\end{align}
where $\hat{\psi}({\mathbf r})$ is the field operator of the molecules in the dressed state $\left\vert+\right\rangle$ and $\mu$ is the chemical potential.

In the superconducting phase, the order parameter in the momentum space takes the form
\begin{align}
\Delta (\mathbf{k})=\int
\frac{d{\mathbf p}}{(2\pi )^3}\widetilde{V}_{\mathrm{eff}}(\mathbf{k}-\mathbf{p})\left\langle c_{-\mathbf{p}}c_{\mathbf{p}}\right\rangle,
\end{align}
where $\widetilde{V}_{\rm eff}({\mathbf k})$ is the Fourier transform of $V_{\rm eff}({\mathbf r})$, $\hat c_{\mathbf{p}}=\int d{\mathbf r}\hat\psi (\mathbf{r})e^{-i\mathbf{p}\cdot{\mathbf r}}/(2\pi)^{3/2}$, and $\left\langle \hat c_{-\mathbf{p}}\hat c_{\mathbf{p}}\right\rangle$ is the pairing function which, within the mean-field theory, reads $\left\langle \hat c_{-\mathbf{p}}\hat c_{\mathbf{p}}\right\rangle =-\Delta (\mathbf{p})\tanh (\beta E_{\mathbf{p}}/2)/(2E_{\mathbf{p}})$. Here, $\beta=1/(k_BT)$ is the inverse temperature and $E_{\mathbf{p}}=\sqrt{\varepsilon _{\mathbf{p}}^{2}+\left\vert \Delta (\mathbf{p})\right\vert ^{2}}$ is the dispersion relation of the Bogoliubov quasiparticle with $\varepsilon _{\mathbf{p}}=p^{2}/(2M)-\mu $. As a result, the gap equation becomes~\cite{Baranov2002}
\begin{align}
\Delta (\mathbf{k})=-\int \frac{d{\mathbf p}}{(2\pi)^{3}}\widetilde V_{\mathrm{eff}}(\mathbf{k}-\mathbf{p})\frac{\tanh (\beta E_{\mathbf{p}}/2)}{2E_{\mathbf{p}}}\Delta (\mathbf{p}).  \label{Delta}
\end{align}
We point out that, unlike an ill-defined potential (e.g., the contact interaction and the pure dipolar interaction) which has to be renormalized~\cite{Baranov2002,Baranov2004,sademelo1993} to ensure the convergence of the integral in Eq.~\eqref{Delta}, the effective potential $V_{\rm eff}$ is well-defined due to its repulsive core. As a result, the two-body wave function behaves as $e^{-B/r^{3}}$ ($B>0$) inside the shielding core, which guarantees that Eq.~\eqref{Delta} is solvable without the interaction renormalization.

To find the critical temperature, we note that when $T$ approaches $T_c$ from below $|\Delta({\mathbf p})|\approx 0$ and, subsequently, $E_{\mathbf p}\approx \varepsilon_{\mathbf p}$. We then expand the gap and interaction in the partial-wave basis as $\Delta (\mathbf{k})=\sum_{lm}Y_{lm}(\hat{k})\Delta _{lm}(k)$ and $\widetilde V_{\mathrm{eff}}(\mathbf{k}-\mathbf{p})=(4\pi )^{2}\sum_{ll^{\prime },m}i^{l^{\prime }-l}Y_{lm}(\hat{k})Y_{l^{\prime }m}^{\ast }(\hat{p})\widetilde V_{ll^{\prime},m}(k,p)$, respectively, where $l$ and $l'$ are odd for identical fermions. Now, Eq.~\eqref{Delta} can be rewritten as
\begin{align}
\Delta_{lm}(k)&=-\frac{2}{\pi }\sum_{l^{\prime}}i^{l^{\prime}-l}\int_{0}^{\infty }p^2dp\widetilde V_{ll^{\prime},m}(k,p)\nonumber\\
&\quad\times \frac{\tanh (\beta_c\varepsilon_{\mathbf{p}}/2)}{2\varepsilon_{\mathbf{p}}}\Delta_{l^{\prime}m}(p), \label{Deltal}
\end{align}
where $\beta_c=(k_BT_c)^{-1}$. Due to the axial symmetry of the system, $\Delta_{lm}$ for different $m$'s are decoupled in Eq.~\eqref{Deltal}. To proceed further, we focus on the weak interaction regime which allows us to approximate the chemical potential by the Fermi energy, i.e., $\mu \approx \varepsilon_{F}=\hbar^2k_F^2/(2M)$. Equation~\eqref{Deltal} then becomes an eigenvalue equation of the integral operator for which negative eigenvalues appears only when $T< T_c$. This condition allows us to find the critical temperature.

Before presenting the results, it is worthwhile to have a brief discussion on the numerical treatment of Eq.~\eqref{Deltal}. For $l=l'=1$, $\widetilde{V}_{ll',m}$ contains a divergent term contributed by the $1/r^6$ shielding potential through the integral $\int_{0}^{\infty}drj_{l}(kr)j_{l^{\prime }}(pr)/r^4$~\cite{SM}, where $j_l$ is the spherical Bessel function. This imposes a difficulty for numerically solving Eq.~\eqref{Deltal}. To remove this divergence, we introduce a truncation $r_{\rm UV}$ on the lower limit of the integration. Consequently, the divergent term is now proportional to $kp/r_{\rm UV}$. Numerically, it is found that the solution of the Eq.~\eqref{Deltal} converges as $r_{\rm UV}$ is much smaller than the size of the shielding core and all results presented below are obtained with the truncation $k_Fr_{\rm UV}\leq 10^{-8}$. This divergence can be understood by considering a two-body problem in the momentum space with interaction potential $\widetilde{V}_{\rm eff}$. Without this divergent term, a false bound state localized inside the shielding core would appear, while the divergent term eliminates the existence of any bound states inside the core. Therefore, the false bound state is removed when $r_{\rm UV}$ is much smaller than the size of the shielding core. This momentum regularization scheme can also be applied to interaction potentials of the form $1/r^{n}$, including the van der Waals and the Lennard-Jones potentials. As to the truncation on angular momentum quantum number, we find that $l_c=9$ is sufficient to ensure the convergence of the solution. It is numerically found that the highest critical temperature is reached when $m=\pm 1$. Without loss of generality, we always take $m=1$ for all results presented below.

\begin{figure}[tbp]
\includegraphics[width=0.95\linewidth]{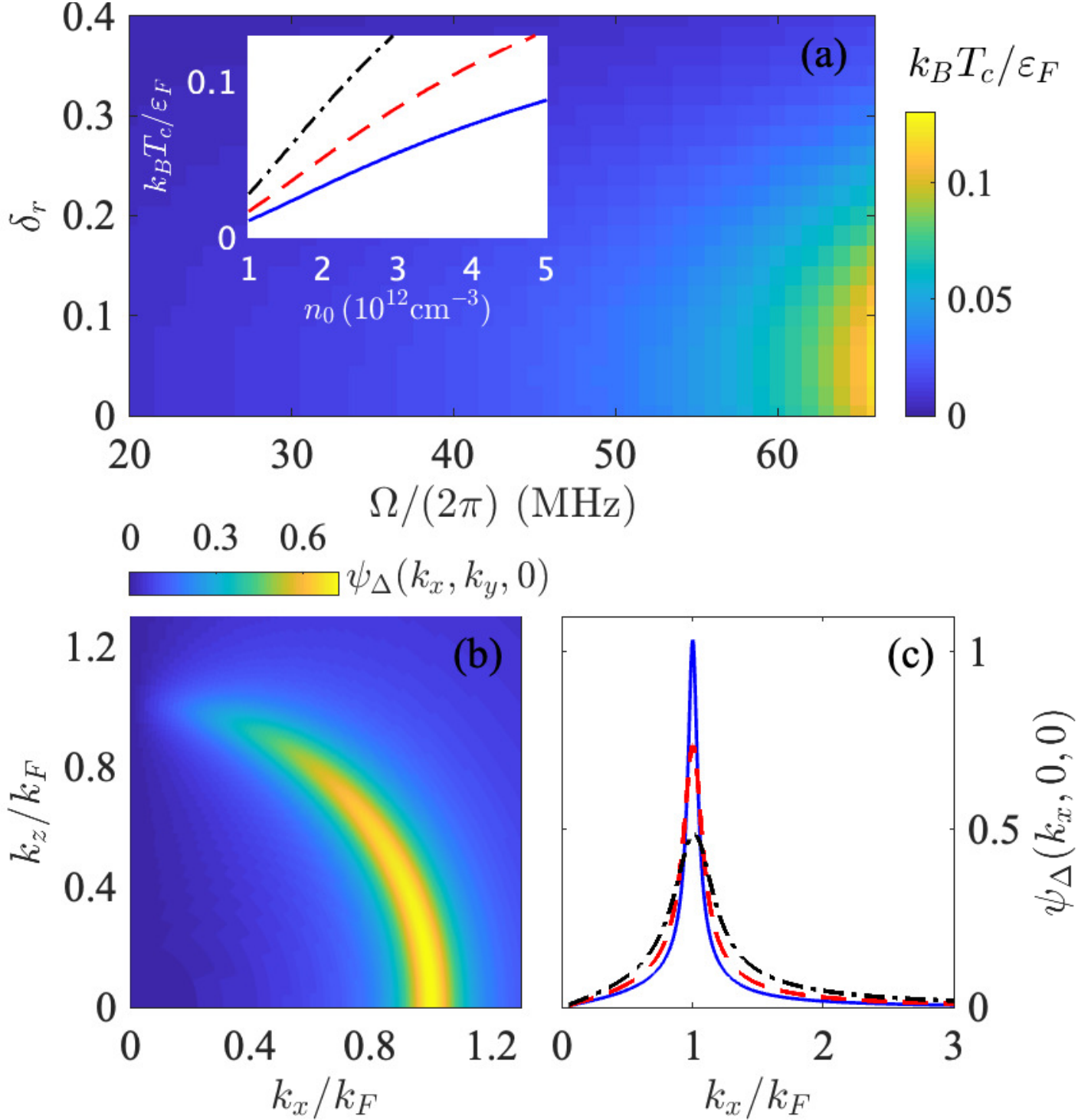}
\caption{(a) Critical temperature as a function of $\Omega$ and $\delta_r$. The inset shows the $n_0$ dependence of $T_c$ for $\delta_r=0.1$ and $\Omega/(2\pi)=28$ (solid line), $38$ (dashed line),  and $48\,\mathrm{MHz}$ (dash-dotted line). (b) Momentum distribution of the normalized pairing function $\psi_{\Delta}(k_x,0,k_z)$ for $\protect\delta =0.1$ and $\Omega/(2\pi)=58\,\mathrm{MHz}$. (c) The normalized pairing function $\psi_{\Delta}(k_x,0,0)$ along the $k_x$ direction for $\delta_r=0.1$ and $\Omega/(2\pi)= 48$ (solid line), $58$ (dashed line), and $66\,{\rm MHz}$ (dash-dotted line). Unless otherwise stated, the density of the gas is $n_{0}=10^{12}\,\mathrm{cm}^{-3}$ in this Figure.}
\label{figCT}
\end{figure}

In Fig.~\ref{figCT}(a), we map out the critical temperature on the $\Omega$-$\delta_r$ plane below the BEC transition temperature $T_c/\varepsilon_F\approx0.137$ in the strong coupling side~\cite{sademelo2006}. Here the control parameters are chosen to avoid the scattering resonances where the weak-interaction assumption is violated. As can be seen, $T_c$ can be increased by either increasing $\Omega$ or decreasing $\delta_r$. This can be roughly understood using Eq.~\eqref{wkbph} as both increasing $\Omega$ and decreasing $\delta_r$ can enhance the attractive interaction. Moreover, the observation that the critical temperature can be efficiently increased by varying $\Omega$ for a fixed effective dipolar strength $C_3$ suggests that the critical temperature is mainly determined by the scattering properties at $k_1\approx k_F$ as Cooper pairs of molecules form at the vicinity of the Fermi surface. In fact, unlike the cross section of the threshold scatterings ($k_1\rightarrow0$) which barely changes for $\Omega/(2\pi)$ up to $80\,{\rm MHz}$ [see Fig.~\ref{figSC}(b)], the scattering cross section at $k_1\approx k_F$ is sensitive to the variation of $\Omega$. The inset of Fig.~\ref{figCT}(a) plots $T_c$ as a function of the gas density $n_0$ for $\delta_{r}=0.1$ and various $\Omega$'s. Apparently, the critical temperature can be dramatically increased with the growing gas density. In particular, for a typical Rabi frequency $\Omega/(2\pi)=38\,\mathrm{MHz}$, $T_{c}$ can be increased by roughly one order of magnitude when $n_{0}$ increases from $10^{12}$ to $5\times 10^{12}\,\mathrm{cm}^{-3}$.

In Fig.~\ref{figCT}(b), we present the momentum distribution of the normalized pairing function $\psi_{\Delta}(\mathbf{k}) \equiv\left\langle c_{\mathbf{k}}c_{-\mathbf{k}}\right\rangle $ on the $k_x$-$k_z$ plane. As can be seen, the pairing function $\psi _{\Delta }(\mathbf{k})$ displays a peak at the Fermi momentum $k_{F}$, suggesting that Cooper pairs are formed around the Fermi surface in the BCS regime. Moreover, $\psi_{\Delta}(\mathbf{k})$ exhibits a clear anisotropy on the $k_x$-$k_z$ plane and is dominated by the $p$ wave as $\int k^2dk\,|\!\int d\hat{\mathbf{k}} \psi_{\Delta}({\mathbf k})Y_{10}(\hat{\mathbf k})|^2 \apprge 0.97$ for all paring functions obtained in this work. Due to the distinct scattering behavior of microwave shielded molecules, this $p$-wave superfluid~\cite{Baranov2002,Baranov2004,Marenko2006,sademelo2006,Jin2008} is achieved over a broad range of Rabi frequency. Finally, we plot, in Fig.~\ref{figCT}(c), $\psi_{\Delta }(\mathbf{k})$ along the $k_x$ direction for $n_{0}=10^{12} \,{\rm cm}^{-3}$, $\delta _{r}=0.1$, and various $\Omega$'s. As can be seen, the width of the peak increases with increasing $\Omega$, indicating that, under a stronger attractive interaction, Cooper pairs are formed over a broader range of momentum.

\textit{Conclusion and discussion}.---We have analytically derived an effective intermolecular potential for  microwave-dressed polar molecules. The validity of the effective potential has been justified by studying the low-energy scatterings of two molecules. It has been shown that the results from single-channel scattering calculations using the effective potential are in good agreement with multi-channel results calculated using the realistic potential. We have also explored the many-body physics of the microwave-dressed NaK gas by calculating the critical temperature of the $p$-wave superfluid and the pairing wave function, which proves that the effective potential is well-behaved and suitable for studying the many-body physics of the molecular gases.

Consider the rich many-body physics across a Feshbach resonance~\cite{sademelo1993,sademelo2006}, it is not surprising to expect similar situation from an ultracold gas of polar molecules in resonant regime. However, as shown in this work, a major challenge for experimental realization of dipolar resonances is the requirement for a large Rabi frequency $\Omega$ that is beyond the reach of the current experiments. It turns out that this can potentially be circumvented by employing an elliptically polarized microwave field~\cite{Luo2022b,SM}. For typical experimental setups, even a small mixing angle can dramatically reduce the resonance Rabi frequency to a value reachable in current experiments.

T.S. thanks Peng Zhang for the stimulating discussion. This work was supported by National Key Research and Development Program of China (Grant No. 2021YFA0718304), by the NSFC (Grants No. 12135018, No. 11974363, No. 12047503, and No. 12274331), and by the Strategic Priority Research Program of CAS (Grant No. XDB28000000).  X.-Y. C. and X.-Y. L. acknowledge support from the Max Planck Society, the European Union (PASQuanS Grant No. 817482) and the Deutsche Forschungsgemeinschaft under German Excellence Strategy -- EXC-2111 -- 390814868 and under Grant No.\ FOR 2247.

\clearpage

\widetext

\begin{center}
\textbf{\large Supplemental Materials}
\end{center}

This Supplemental Material is structured as follows. In Section~\ref{sm:PS}, we derive the effective potential $V_{\mathrm{eff}}$ for two molecules in the highest dressed state channel. In Section~\ref{TBS}, we first outline formulation for the multichannel scattering calculation and then present additional comparison to justify the effective potential. In Section~\ref{TI}, we derive the partial-wave expansion of the effective potential in the momentum space.

\setcounter{equation}{0} \setcounter{figure}{0} \setcounter{table}{0} %
\setcounter{page}{1} \setcounter{section}{0} \makeatletter
\renewcommand{\theequation}{S\arabic{equation}} \renewcommand{\thefigure}{S%
\arabic{figure}} \renewcommand{\bibnumfmt}[1]{[S#1]} \renewcommand{%
\citenumfont}[1]{S#1} \renewcommand{\thesection}{S\arabic{section}}%
\setcounter{secnumdepth}{3}

\section{Derivation of the effective potential} \label{sm:PS}
To obtain the effective interaction potential between two molecules in the highest dressed-state channel. For convenience, let us first write out the two-particle Hamiltonian
\begin{align}
\hat H_2\approx\sum_{j=1,2}\hat h_{\rm in}(j)+V({\mathbf r}_1-{\mathbf r}_2)
\end{align}
where we have neglected the kinetic energy via the Born-Oppenheimer approximation. Moreover, it is more convenient to express the dipole-dipole interaction as
\begin{equation}
V({\mathbf r})=-8\sqrt{\frac{2}{15}}\pi ^{3/2}\frac{d^{2}}{4\pi \epsilon_{0}r^{3}}%
\sum_{m=-2}^{2}Y_{2m}^{\ast}(\hat{\mathbf r})\Sigma_{2,m},
\end{equation}%
where $Y_{2m}(\hat{\mathbf r})$ are spherical harmonics and $\Sigma_2$ is an rank-2 spherical tensor with components
\begin{align}
\Sigma _{2,0} =\frac{1}{\sqrt{6}}(\hat{d}_{1}^{+}\hat{d}_{2}^{-} +\hat{d}_{1}^{-}\hat{d}_{2}^{+}+2\hat{d}_{1}^{0}\hat{d}_{2}^{0}),\;
\Sigma _{2,\pm 1}=\frac{1}{\sqrt{2}}(\hat{d}_{1}^{\pm }\hat{d}_{2}^{0}+\hat{d}_{1}^{0}\hat{d}_{2}^{\pm }),\;\mbox{and }\Sigma _{2,\pm 2}=\hat{d}_{1}^{\pm }\hat{d}_{2}^{\pm }.\label{Sigma}
\end{align}
Here $\hat{d}_{j}^{\pm }=Y_{1,\pm 1}(\mathbf{\hat{d}}_{j})$ and
$\hat{d}_{j}^{0}=Y_{1,0}(\mathbf{\hat{d}}_{j})$ with $\mathbf{\hat{d}}_{j=1,2}$ being the unit vector along the directions of the $j$th dipole moment. Apparently, we have $\hat{d}_{j}^{-}=-(\hat{d}_{j}^{+})^{\dagger }$.

In principle, to obtain an effective potential, one should diagonalize $\hat{H}_2$ in the two-particle Hilbert space $\mathcal{H}_1\otimes\mathcal{H}_1$, where $\mathcal{H}_1={\rm span}\{|0,0\rangle,|1,1\rangle,|1,0\rangle,|1,-1\rangle\}$ is the Hilbert space for the internal states of a molecule. To this end, we  first note that, in the frame co-rotating with the microwave, the operators $\hat{d}^{\pm ,0}$ can be expressed as
\begin{align}
\hat{d}^{0} =\frac{1}{\sqrt{4\pi }}(\left\vert 1,0\right\rangle
\left\langle 0,0\right\vert e^{i\omega _{0}t}+\mathrm{H.c.}),  \;
\hat{d}^{+} =\frac{1}{\sqrt{4\pi }}(\left\vert 1,1\right\rangle
\left\langle 0,0\right\vert e^{i\omega _{0}t}-\left\vert 0,0\right\rangle
\left\langle 1,-1\right\vert e^{-i\omega _{0}t}),\mbox{ and }\hat{d}^{-}=-(\hat{d}^{+})^\dag.\label{spherd}
\end{align}%
After substituting Eq.~\eqref{spherd} into \eqref{Sigma}, the components of the rank-2 spherical tensor $\Sigma_2$ become
\begin{align}
\Sigma _{2,0} &=\frac{1}{4\pi \sqrt{6}}\Big(2\left\vert 1,0\right\rangle
\left\langle 0,0\right\vert \otimes \left\vert 0,0\right\rangle \left\langle
1,0\right\vert -\left\vert 1,1\right\rangle \left\langle 0,0\right\vert
\otimes \left\vert 0,0\right\rangle \left\langle 1,1\right\vert -\left\vert
0,0\right\rangle \left\langle 1,-1\right\vert \otimes \left\vert
1,-1\right\rangle \left\langle 0,0\right\vert +\mathrm{H.c.}\Big)  \notag \\
\Sigma _{2,1} &=\frac{1}{4\pi \sqrt{2}}\Big(\left\vert 1,1\right\rangle
\left\langle 0,0\right\vert \otimes \left\vert 0,0\right\rangle \left\langle
1,0\right\vert -\left\vert 0,0\right\rangle \left\langle 1,-1\right\vert
\otimes \left\vert 1,0\right\rangle \left\langle 0,0\right\vert +\left\vert
0,0\right\rangle \left\langle 1,0\right\vert \otimes \left\vert
1,1\right\rangle \left\langle 0,0\right\vert \nonumber\\
&\quad-\left\vert 1,0\right\rangle
\left\langle 0,0\right\vert \otimes \left\vert 0,0\right\rangle \left\langle
1,-1\right\vert \Big),  \notag \\
\Sigma_{2,-1}&=-\Sigma _{2,1}^{\dagger },\nonumber\\
\Sigma _{2,2} &=-\frac{1}{4\pi }\Big(\left\vert 1,1\right\rangle \left\langle
0,0\right\vert \otimes \left\vert 0,0\right\rangle \left\langle
1,-1\right\vert +\left\vert 0,0\right\rangle \left\langle 1,-1\right\vert
\otimes \left\vert 1,1\right\rangle \left\langle 0,0\right\vert \Big),\nonumber\\
\Sigma _{2,-2}&=\Sigma_{2,2}^{\dagger },\nonumber
\end{align}%
where we have adopted the rotating-wave approximation by neglecting the high-frequency oscillation terms.

\begin{figure}[tbp]
\includegraphics[width=0.7\linewidth]{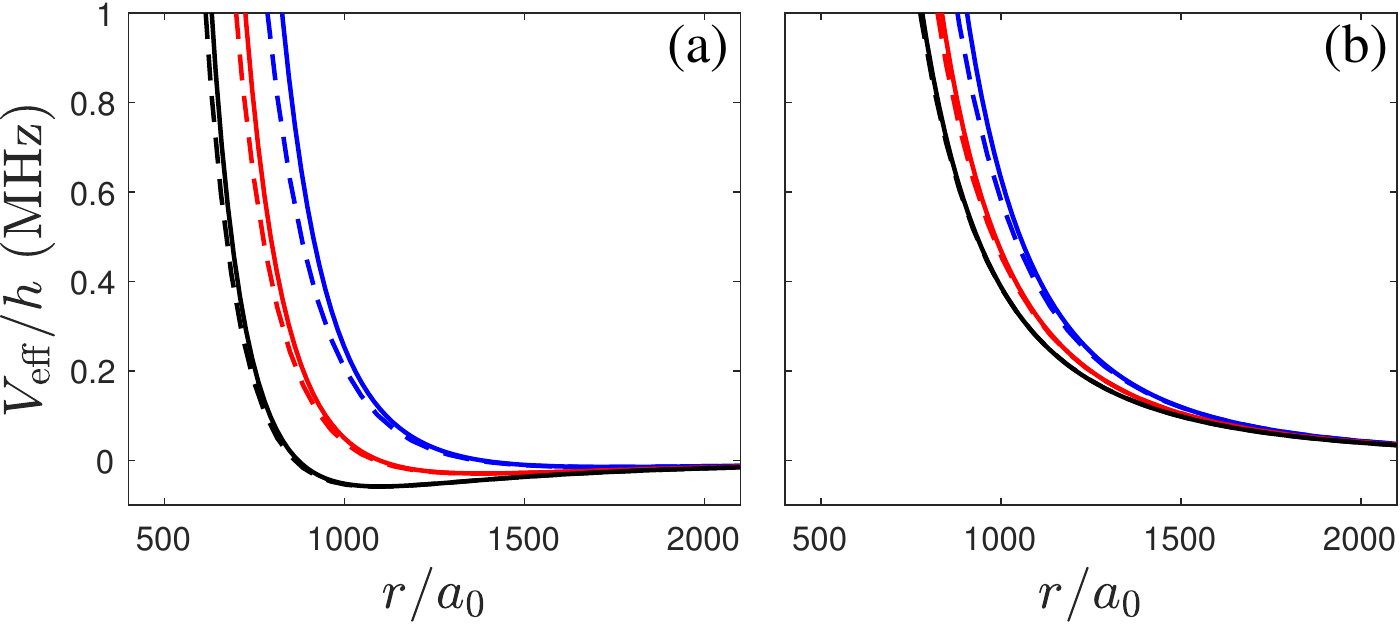}
\caption{Comparison of the exact adiabatic potential (solid lines) with the fitted effective potential (dashed line) along the angles $\protect\theta=\protect\pi/3$ (a) and $\protect\pi/4$ (b). Other parameters are $\protect\delta_r=0.1$ and $\Omega/(2\protect\pi)=20$, $50$, and $80\,\mathrm{%
MHz}$ (for three sets of curves in descending order).}
\label{SM1}
\end{figure}

To diagonalize $\hat H_2$, it is more convenient to proceed in the dressed-state basis $\{|+\rangle,|0\rangle,|-1\rangle,|-\rangle\}$ in which $\hat h_{\rm in}(j)$ is diagonal. Moreover, because $\hat H_2$ possesses a parity symmetry, its eigenstates must be either symmetric or antisymmetric. The fact that the microwave shielded state lies in the symmetric sector allows us to focus on the ten-dimensional symmetric subspace. Fortunately, it turns out that $V({\mathbf r})$ in the seven-dimensional symmetric subspace, $\mathcal{S}_7$, is decoupled from the remaining three-dimensional symmetric subspace. Then under the basis $\{|\nu\rangle\}_{\nu=1}^7$, we have
\begin{align}
\hat h_{\rm in}(1)+\hat h_{\rm in}(2)&=\begin{pmatrix}
0&0&0&0&0&0&0\\
0&\frac{1}{2}(\delta-\Omega_{\mathrm{eff}})&0&0&0&0&0\\
0&0&\frac{1}{2}(\delta-\Omega_{\mathrm{eff}})&0&0&0&0\\
0&0&0&-\Omega_{\mathrm{eff}}&0&0&0\\
0&0&0&0&\frac{1}{2}(\delta-3\Omega_{\mathrm{eff}})&0&0\\
0&0&0&0&0&\frac{1}{2}(\delta-3\Omega_{\mathrm{eff}})&0\\
0&0&0&0&0&0&-2\Omega_{\mathrm{eff}}\\
\end{pmatrix},\nonumber\\
\Sigma _{2,0} &=\frac{1}{4\pi \sqrt{6}}\begin{pmatrix}
-2u^{2}v^{2} & 0 & 0 & -\sqrt{2}uv(u^{2}-v^{2}) & 0 & 0 & 2u^{2}v^{2} \\
0 & 2u^{2} & 0 & 0 & -2uv & 0 & 0 \\
0 & 0 & -u^{2} & 0 & 0 & uv & 0 \\
-\sqrt{2}uv(u^{2}-v^{2}) & 0 & 0 & -(u^{2}-v^{2})^{2} & 0 & 0 & \sqrt{2}%
uv(u^{2}-v^{2}) \\
0 & -2uv & 0 & 0 & 2v^{2} & 0 & 0 \\
0 & 0 & uv & 0 & 0 & -v^{2} & 0 \\
2u^{2}v^{2} & 0 & 0 & \sqrt{2}uv(u^{2}-v^{2}) & 0 & 0 & -2u^{2}v^{2}%
\end{pmatrix},  \notag \\
\Sigma _{2,1} &=\frac{1}{4\pi \sqrt{2}}\begin{pmatrix}
0 & \sqrt{2}u^{2}v & 0 & 0 & -\sqrt{2}uv^{2} & 0 & 0 \\
0 & 0 & -u^{2} & 0 & 0 & uv & 0 \\
0 & 0 & 0 & 0 & 0 & 0 & 0 \\
0 & u(u^{2}-v^{2}) & 0 & 0 & -v(u^{2}-v^{2}) & 0 & 0 \\
0 & 0 & uv & 0 & 0 & -v^{2} & 0 \\
0 & 0 & 0 & 0 & 0 & 0 & 0 \\
0 & -\sqrt{2}u^{2}v & 0 & 0 & \sqrt{2}uv^{2} & 0 & 0%
\end{pmatrix},  \notag \\
\Sigma _{2,2} &=-\frac{1}{4\pi}\begin{pmatrix}
0 & 0 & \sqrt{2}u^{2}v & 0 & 0 & -\sqrt{2}uv^{2} & 0 \\
0 & 0 & 0 & 0 & 0 & 0 & 0 \\
0 & 0 & 0 & 0 & 0 & 0 & 0 \\
0 & 0 & u(u^{2}-v^{2}) & 0 & 0 & -v(u^{2}-v^{2}) & 0 \\
0 & 0 & 0 & 0 & 0 & 0 & 0 \\
0 & 0 & 0 & 0 & 0 & 0 & 0 \\
0 & 0 & -\sqrt{2}u^{2}v & 0 & 0 & \sqrt{2}uv^{2} & 0%
\end{pmatrix}.\nonumber
\end{align}
Now, for a given $\mathbf{r}$, we diagonal the $7\times7$ matrix
\begin{align}
\hat H_2=\sum_{j=1,2}\hat h_{\rm in}(j)-8\sqrt{\frac{2}{15}}\pi ^{3/2}\frac{d^{2}}{4\pi \epsilon_{0}r^{3}}%
\sum_{m=-2}^{2}Y_{2m}^{\ast}(\hat{\mathbf r})\Sigma_{2,m},
\end{align}
the highest adiabatic curve then corresponds to the effective potential for the microwave-shielded state. Interestingly, it is found that, in the region where the inter-molecular distance is larger than the shielding core, the contributions to the highest eigenstate is dominated by the states $\left\vert1\right\rangle $, $\left\vert 2\right\rangle$, and $|3\rangle$. Therefore, we may project $\hat H_2$ onto the three-dimemsional subspace spanned by  $\{|1\rangle,|2\rangle,|3\rangle\}$ and obtain the reduced two-body Hamiltonian
\begin{equation}
\hat H_{2}'(\mathbf{r})=
\begin{pmatrix}
Du^{2}v^{2}(\cos ^{2}\theta -\frac{1}{3}) & Du^{2}v\sin \theta \cos \theta e^{-i\varphi } & \frac{1}{\sqrt{2}}D
u^{2}v\sin ^{2}\theta e^{-2i\varphi } \\
Du^{2}v\sin \theta \cos \theta e^{i\varphi } & \frac{1}{2}(\delta -\Omega _{\mathrm{eff}})-Du^{2}(\cos ^{2}\theta -\frac{1}{3}) & -\frac{1}{\sqrt{2}}Du^{2}\sin \theta \cos \theta e^{-i\varphi } \\
\frac{1}{\sqrt{2}}Du^{2}v\sin ^{2}\theta e^{2i\varphi } & -\frac{1}{\sqrt{2}} Du^{2}\sin \theta \cos \theta e^{i\varphi } & \frac{1}{2}(\delta -\Omega_{\mathrm{eff}})+\frac{1}{2}Du^{2}(\cos^{2}\theta -\frac{1}{3})%
\end{pmatrix}.
\end{equation}%
where $D=d^{2}/(4\pi \epsilon _{0}r^{3})$. We then introduce a unitary transformation
\begin{equation}
U_{2}=\begin{pmatrix}
1 & 0 & 0 \\
0 & \sqrt{\frac{2}{\cos ^{2}\theta +1}}\cos \theta  & -\frac{1}{\sqrt{\cos
^{2}\theta +1}}\sin \theta e^{-i\varphi } \\
0 & \frac{1}{\sqrt{\cos ^{2}\theta +1}}\sin \theta e^{i\varphi } & \sqrt{%
\frac{2}{\cos ^{2}\theta +1}}\cos \theta
\end{pmatrix}
\end{equation}
which transfers $\hat H_2'$ into
\begin{equation}
\hat H_2''\equiv U_{2}^{\dagger}\hat H_{2}'U_{2}=\begin{pmatrix}
Du^{2}v^{2}(\cos ^{2}\theta -\frac{1}{3}) & \frac{1}{\sqrt{2}}Du^{2}v\sin \theta \sqrt{\cos ^{2}\theta +1}e^{-i\varphi } & 0 \\
\frac{1}{\sqrt{2}}Du^{2}v\sin \theta
\sqrt{\cos ^{2}\theta +1}e^{i\varphi } & \frac{1}{2}(\delta -\Omega _{%
\mathrm{eff}})-\frac{1}{2}Du^2(\cos
^{2}\theta +\frac{1}{3}) & 0 \\
0 & 0 & \frac{1}{2}(\delta -\Omega _{\mathrm{eff}})+\frac{1}{3}Du^{2}%
\end{pmatrix}.
\end{equation}%
We can now diagonalize the upper $2\times 2$ matrix and obtain the adiabatic potential curve of the highest dressed-state channel $|1\rangle$,
\begin{align}
V_{\mathrm{adia}}^{|1\rangle}(\mathbf{r}) &=\frac{1}{2}\left[\frac{1}{2}(\delta -\Omega _{%
\mathrm{eff}})+\frac{1}{3}Du^{2}v^{2}(3\cos^{2}\theta -1)-\frac{1}{6}Du^{2}(3\cos ^{2}\theta +1)\right] \nonumber\\
&\quad+\sqrt{\frac{1}{4}\left[-\frac{1}{2}(\delta -\Omega _{\mathrm{eff}})+\frac{1}{3}Du^{2}v^{2}(3\cos ^{2}\theta -1)+
\frac{1}{6}Du^2(3\cos ^{2}\theta +1)\right]^{2}+\frac{1}{2}D^{2}u^{4}v^{2}\sin ^{2}\theta (\cos ^{2}\theta +1)}.
\end{align}
After expanding $V_{\mathrm{adia}}^{|1\rangle}(\mathbf{r})$ to the order $D^2$, we obtain the effective interaction potential $V_{\mathrm{eff}}(\mathbf{r})$. It should be noted that the effective potential works very well when the intermolecular distance satisfies $r^{3}>d^{2}/(4\pi\epsilon _{0}\Omega )$. For the validity of the effective potential, we present addition comparisons for $\theta =\pi /3$ and $\pi /4$ in Fig. \ref{SM1}. As can be seen, the effective potentials $V_{\mathrm{eff}}(\mathbf{r})$ and the exact adiabatic potential are in very good agreement.

Finally, in addition to the circular polarized microwave as discussed in the present, the elliptically polarized microwave field of the form $\Omega (\cos \alpha_{m}\sigma^{+}+\sin \alpha_{m}\sigma^{-})$ can also be used to shield the molecules, where $\alpha_m$ is the elliptic angle of the left- and right-circularized microwave fields. Then, for small elliptic angle $\alpha_m$, one can also derive an analytic expression for the effective potential by following the same procedure, i.e.,
\begin{align}
V_{\mathrm{eff}}(\alpha_m,\mathbf{r}) &=\frac{C_{3}}{2r^{3}}\left(3\cos
^{2}\theta -1+3\sin 2\alpha_m\sin ^{2}\theta \cos 2\varphi \right)
\notag \\
&\quad+\frac{35C_{6}}{4r^{6}}\sin ^{2}\theta \left(\cos ^{2}\theta +1-2\sin
2\alpha_m\cos ^{2}\theta \cos 2\varphi-\sin ^{2}2\alpha_m\sin ^{2}\theta \cos ^{2}2\varphi \right).
\label{Vpe}
\end{align}

\section{Two-body scatterings} \label{TBS}
Here we outline the procedure for solving the multi-channel scattering problem and present additional comparison between the single- and multi-channel results. To this end, let us first write down the multi-channel Schr\"odinger equations
\begin{align}
\sum_{\nu^{\prime}=1}^7\left(-\frac{\hbar^{2}\nabla ^{2}}{M}\delta _{\nu \nu^{\prime}}+V_{\nu \nu^{\prime }}\right)\psi _{\nu ^{\prime }}({\mathbf{r}})=\frac{\hbar^2k_{\nu}^2}{M}\psi_{\nu}({\mathbf{r}}),\label{smscase}
\end{align}
where $\hbar^2k_{\nu}^{2}/M$ is the incident energy of the $\nu$th channel. In general, one should expand $\psi_{\nu}$ in the partial wave basis as $\psi_{\nu}=\sum_{l_\nu m_\nu}r^{-1}\phi_{\nu l_\nu m_\nu}Y_{l_\nu m_\nu}(\hat{\mathbf r})$ with $l_\nu$ being odd. However, it turns out that, for a given $m$, only these partial wave with $m_{\nu}=m,m+1,m+2,m,m+1,m+2$, and $m$ for $\nu=1$ to $7$ are coupled. Therefore, we may treat each $m$ separately by expanding the scattering wave function as
\begin{eqnarray}
\psi _{\nu =1,4,7}(\mathbf{r}) &=&\sum_{l}\frac{1}{r}Y_{lm}(\hat{r})\phi
_{\nu l}(r),  \notag \\
\psi _{\nu =2,5}(\mathbf{r}) &=&\sum_{l}\frac{1}{r}Y_{lm+1}(\hat{r})\phi
_{\nu l}(r),  \notag \\
\psi _{\nu =3,6}(\mathbf{r}) &=&\sum_{l}\frac{1}{r}Y_{lm+2}(\hat{r})\phi
_{\nu l}(r),\nonumber
\end{eqnarray}%
where we have ignored the quantum number for the projection of the orbital angular momentum. Now, the Schr\"odinger for the partial waves becomes
\begin{equation}
\sum_{\nu ^{\prime }l^{\prime }}\left[-\frac{\hbar^2}{M}\left(\frac{\partial^2}{\partial r^2}-\frac{l(l+1)%
}{r^{2}}\right)\delta _{\nu \nu ^{\prime }}\delta _{ll^{\prime }}-\frac{d_{0}^{2}}{%
4\pi \epsilon _{0}r^{3}}\sum_{j}(\Xi _{j}^{m})_{\nu l,\nu ^{\prime
}l^{\prime }}\right]\phi _{\nu ^{\prime }l^{\prime }}(r)=\frac{k_{\nu }^{2}}{M}%
\phi _{\nu l}(r),  \label{SECC}
\end{equation}%
where
\begin{eqnarray}
(\Xi _{0}^{m})_{\nu l,\nu ^{\prime }l^{\prime }} &=&4\pi \sqrt{\frac{%
2(2l^{\prime }+1)}{3(2l+1)}}\Sigma _{2,0}C_{l^{\prime
}020}^{l0}\,\mathrm{diag}(C_{l^{\prime }m20}^{lm},C_{l^{\prime
}m+120}^{lm+1},C_{l^{\prime }m+220}^{lm+2},C_{l^{\prime
}m20}^{lm},C_{l^{\prime }m+120}^{lm+1},C_{l^{\prime
}m+220}^{lm+2},C_{l^{\prime }m20}^{lm}),  \notag \\
(\Xi _{1}^{m})_{\nu l,\nu ^{\prime }l^{\prime }} &=&-4\pi \sqrt{\frac{%
2(2l^{\prime }+1)}{3(2l+1)}}\Sigma _{2,1}C_{l^{\prime
}020}^{l0}\,\mathrm{diag}(0,C_{l^{\prime }m+12-1}^{lm},C_{l^{\prime
}m+22-1}^{lm+1},0,C_{l^{\prime }m+12-1}^{lm},C_{l^{\prime }m+22-1}^{lm+1},0),
\notag \\
(\Xi _{2}^{m})_{\nu l,\nu ^{\prime }l^{\prime }} &=&4\pi \sqrt{\frac{%
2(2l^{\prime }+1)}{3(2l+1)}}\Sigma _{2,2}C_{l^{\prime }020}^{l0}C_{l^{\prime
}m+22-2}^{lm},\nonumber
\end{eqnarray}%
are the interaction matrix elements with $\mathrm{diag}(\cdots)$ denoting the diagonal matrix and $C_{l^{\prime }m^{\prime}l''m''}^{lm}\equiv \langle lm|l'm';l''m''\rangle$ the short-hand notation for the Clebsch-Gordan coefficients. Apparently, we have $\Xi _{-1}^{m}=\Xi _{1}^{m\dagger }$ and $\Xi _{-2}^{m}=\Xi _{2}^{m\dagger }$.

The coupled channel Schr\"{o}dinger Eq. (\ref{SECC}) can be solved using the log-derivative method with high precision. In the compact form,
Eq. (\ref{SECC}) reads%
\begin{equation}
\left[ \frac{\partial^2}{\partial r^2}+\mathcal{V}(r)\right]\phi (r)=0,  \label{cSE}
\end{equation}%
where the potential%
\begin{equation}
\mathcal{V}(r)=\left[k_{\nu }^{2}-\frac{l(l+1)}{r^{2}}\right]\delta _{\nu \nu ^{\prime
}}\delta _{ll^{\prime }}+M\frac{d_{0}^{2}}{4\pi \epsilon _{0}r^{3}}%
\sum_{j}(\Xi _{j}^{m})_{\nu l,\nu ^{\prime }l^{\prime }}.
\end{equation}%
Now we define the matrix $\mathbf{\phi }(r)$ whose columns denote the
scattering wavefunctions for the incident waves in different channels, i.e.,
the angular momentum $l$ and dressed-state channels. It follows from Eq. (%
\ref{cSE}) that the matrix $\mathcal{Y}(r)=\partial _{r}\mathbf{\phi }(r)%
\mathbf{\phi }^{-1}(r)$ satisfies%
\begin{equation}
\partial _{r}\mathcal{Y}(r)=-\mathcal{V}(r)-\mathcal{Y}^{2}(r).  \label{Y}
\end{equation}%
By solving Eq. (\ref{Y}) numerically with an appropriate boundary condition $%
\mathcal{Y}(0)=10^{20}$, we obtain $\mathcal{Y}(r)$ in the asymptotic limit $%
r\rightarrow \infty $.

On the other hand, for the incident particles with angular momentum $l$ in
the dressed-state channel $\nu \in S_{7}$, the asymptotic wavefunction ($%
r\rightarrow \infty $)%
\begin{equation}
\phi _{\nu ^{\prime }l^{\prime },\nu l}^{\mathrm{a}}(r)=\mathbf{J}%
_{l}(k_{\nu }r)\delta _{ll^{\prime }}\delta _{\nu \nu ^{\prime }}+\sum_{\nu
^{\prime }l^{\prime }}\mathbf{N}_{l^{\prime }}(k_{\nu ^{\prime }}r)K_{\nu
^{\prime }l^{\prime },\nu l},  \label{pa}
\end{equation}%
can be obtained analytically, where $\mathbf{J}_{l}(k_{\nu }r)=k_{\nu
}^{-1/2}\hat{\jmath}_{l}(k_{\nu }r)$ and $\mathbf{N}_{l}(k_{\nu }r)=k_{\nu
}^{-1/2}\hat{n}_{l}(k_{\nu }r)$ are determined by the Riccati-Bessel
functions $\hat{\jmath}_{l}(z)=zj_{l}(z)$ and $\hat{n}_{l}(z)=-zn_{l}(z)$,
and $K_{\nu ^{\prime }l^{\prime },\nu l}$ is the $K$-matrix element for
in-coming particles with angular momentum $l$ in the channel $\nu $ and
out-going particles with angular momentum $l^{\prime }$ in the channel $\nu
^{\prime }$. By matching the numerical solution $\mathcal{Y}(r\rightarrow
\infty )$ and $\partial _{r}\mathbf{\phi }^{\mathrm{a}}(r)\mathbf{\phi }^{%
\mathrm{a}-1}(r)$ determined by Eq. (\ref{pa}), we can obtain the $K$-matrix
element and the scattering amplitude%
\begin{equation}
f_{\nu ^{\prime }l^{\prime },\nu l}=i\frac{1}{\sqrt{k_{\nu ^{\prime }}}}\left(
\frac{1}{K+i}K\right)_{\nu ^{\prime }l^{\prime },\nu l}\frac{1}{\sqrt{k_{\nu }}}
\end{equation}%
from the channel $(\nu l)$ to the channel $(\nu ^{\prime }l^{\prime })$, and
the partial wave scattering cross section $\sigma _{\nu ^{\prime }l^{\prime
},\nu l}=4\pi \left\vert f_{\nu ^{\prime }l^{\prime },\nu l}\right\vert ^{2}$%
. We can also define the average elastic and inelastic scattering cross
sections $\sigma _{\nu }^{\mathrm{el}}=\sum_{ll^{\prime }}\sigma _{\nu
l^{\prime },\nu l}$ and $\sigma _{\nu }^{\mathrm{inel}}=\sum_{\nu ^{\prime
},ll^{\prime }}k_{\nu ^{\prime }}\sigma _{\nu ^{\prime }l^{\prime },\nu
l}/k_{\nu }$.

To justify the effective potential, we solve the single channel Schr\"{o}%
dinger equation%
\begin{equation}
\left[ -\frac{\hbar^2\nabla^{2}}{M}+V_{\mathrm{eff}}(\mathbf{r})\right]\psi _{++}(%
\mathbf{r})=\frac{k_{1}^{2}}{M}\psi _{++}(\mathbf{r})
\end{equation}%
with the effective potential $V_{\mathrm{eff}}(\mathbf{r})$. In the angular
momentum basis $\psi _{++}(\mathbf{r})=\sum_{l}Y_{lm}(\hat{r})\phi _{l}(r)/r$%
, $\phi _{l}(r)$ satisfies%
\begin{equation}
\sum_{l^{\prime }}\left[-\frac{\hbar^2}{M}\left(\frac{\partial^2}{\partial {r}^{2}}-\frac{l(l+1)}{r^{2}}%
\right)\delta _{ll^{\prime }}+V_{\mathrm{eff,}ll^{\prime }}^{m}(r)\right]\phi
_{l^{\prime }}(r)=\frac{k_{1}^{2}}{M}\phi _{l}(r),  \label{SECl}
\end{equation}%
where the pseudo-potential%
\begin{eqnarray}
V_{\mathrm{eff,}ll^{\prime }}^{m}(r) =\frac{C_{3}}{r^{3}}\sqrt{\frac{%
2l^{\prime }+1}{2l+1}}C_{l^{\prime }020}^{l0}C_{l^{\prime }m20}^{lm}+\frac{%
C_{6}}{r^{6}}\sqrt{\frac{2l^{\prime }+1}{2l+1}}(7\delta _{ll^{\prime }}-5C_{l^{\prime }020}^{l0}C_{l^{\prime
}m20}^{lm}-2C_{l^{\prime }040}^{l0}C_{l^{\prime }m40}^{lm}).
\end{eqnarray}%
The Eq. (\ref{SECl}) is solved via the log-derivative method. By defining%
\begin{equation}
\mathcal{V}(r)=[k_{1}^{2}-\frac{l(l+1)}{r^{2}}]\delta _{ll^{\prime }}-MV_{%
\mathrm{eff,}ll^{\prime }}^{m}(r),  \label{Vr}
\end{equation}%
we can rewrite Eq. (\ref{SECl}) as the same form of Eq. (\ref{cSE}) and
obtain $\mathcal{Y}(r)$. By matching the numerical solution $\mathcal{Y}%
(r\rightarrow \infty )$ with the analytical asymptotic wavefunction%
\begin{equation}
\phi _{l^{\prime }l}^{\mathrm{a}}(r)=\mathbf{J}_{l}(k_{1}r)\delta
_{ll^{\prime }}+\sum_{l^{\prime }}\mathbf{N}_{l^{\prime
}}(k_{1}r)K_{l^{\prime }l}
\end{equation}%
in the single dressed state channel $\nu =1$, we obtain the $K$-matrix and
the scattering amplitude $f_{l^{\prime }l}=i(\frac{1}{K+i}K)_{l^{\prime
}l}/k_{1}$, which results in the scattering amplitude $\sigma _{l^{\prime
}l}=4\pi \left\vert f_{l^{\prime }l}\right\vert ^{2}$.

\begin{figure}[tbp]
\includegraphics[width=0.75\linewidth]{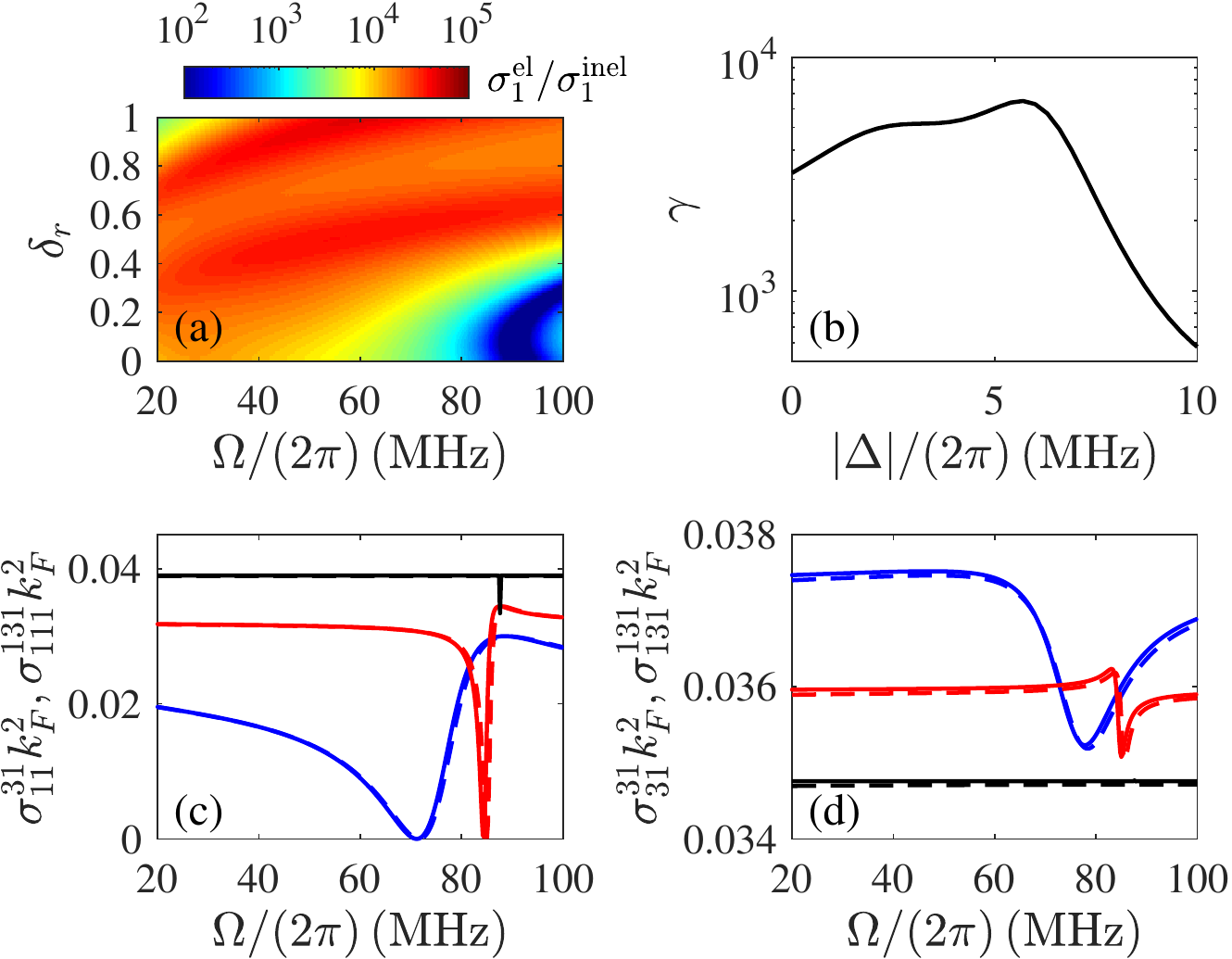}
\caption{(a) The ratio $\protect\sigma _{1}^{\mathrm{el}}/\protect\sigma %
_{1}^{\mathrm{inel}}$ for the incident momentum $k_{1}/k_{F}=0.45$. (b) $\protect\gamma$-ratio as a function of the detuning for $T=800\,{\rm nK}$ and $\Omega/(2\protect\pi)=11\,{\rm MHz}$. (c) The scattering cross sections $\protect\sigma_{11}^{31}k_F^2$ (solid lines) and $\protect\sigma_{111}^{131}k_F^2$ (dashed lines) as functions of $\Omega$ for
$\protect\delta_r=0.1$ and $k_{1}/k_{F}=0.04$ (black lines), $0.45$ (red
lines), and $1$ (blue lines). (d) The scattering cross sections $\protect%
\sigma_{31}^{31}k_F^2$ (solid lines) and $\protect\sigma_{131}^{131}k_F^2$
(dashed lines) as functions of $\Omega$ for $\protect\delta_r=0.1$ and $%
k_{1}/k_{F}=0.04$ (black lines), $0.45$ (red lines), and $1$ (blue lines).}
\label{SM2}
\end{figure}

In Fig.~\ref{SM2}, we present additional results on the scattering properties. Figure~\ref{SM2}(a) for the incident momentum $k_{1}/k_{F}=0.45$, $\sigma_{1}^{\mathrm{el}}/\sigma _{1}^{\mathrm{inel}}$ shows the small inelastic
scattering ($\sigma _{1}^{\mathrm{el}}/\sigma _{1}^{\mathrm{inel}}>10^{3}$) away from the shape resonance (the blue region), which guarantees the projection in the single dressed state channel $\nu =1$. In Fig. \ref{SM2}(b), we plot the $\gamma $-ratio~[30] of elastic to inelastic collision rates as a function of $\Delta $ for $\Omega /(2\pi )=11\,{\rm MHz}$, which agrees with the experiment very well. To validate the effective potential $V_{\mathrm{eff}}(\mathbf{r})$, we have shown the good agreement of $\sigma _{11}^{11}$ and $\sigma _{111}^{111}$ in the $p$-wave channel in Fig. 2 of the main text. Here, in Fig. \ref{SM2}(c) and (d), we compare the scattering cross sections $\sigma _{11}^{31}$ with $\sigma_{111}^{131}$ and $\sigma_{31}^{31}$ with $\sigma_{131}^{131}$, respectively. Both exhibit good agreement, which justifies the applicability of $V_{\mathrm{eff}}(\mathbf{r})$.

We also perform calculations using the multi-channel potential and the effective potential~\ref{Vpe} for the microwave field with the elliptic angle $\alpha_{m}$, which agree with each other very well for $\alpha_{m}<20^\circ$. Remarkably, for a typical experimental setup, even a small elliptic angle, say $\alpha_m=5^\circ$, can be dramatically reduce the resonance Rabi frequency to about $22\,{\rm MHz}$, a value within the reach of the current experimental technique.

\section{Effective potential in the momentum space} \label{TI}
Here we derive the partial wave expansion of the effective potential in the momentum space. To this end, we first note that the Fourier transform of the effective potential is
\begin{equation}
\widetilde{V}_{\mathrm{eff}}(\mathbf{k}-\mathbf{p})=\int d^{3}re^{-i\mathbf{%
k\cdot r}}V_{\mathrm{eff}}(\mathbf{r})e^{i\mathbf{p\cdot r}}.
\end{equation}%
Making use of the partial wave expansion for the plane wave $e^{i\mathbf{k\cdot r}}=4\pi \sum_{lm}i^{l}j_{l}(kr)Y_{lm}(\hat{r})Y_{lm}^{\ast }(\hat{k})$, the effective potential can be expanded as
\begin{align}
\tilde{V}_{\mathrm{eff}
}(\mathbf{k}-\mathbf{p})=(4\pi )^{2}\sum_{ll^{\prime }m}i^{l^{\prime
}-l}Y_{lm}(\hat{k})Y_{l^{\prime }m}^{\ast }(\hat{p})\tilde{V}_{ll^{\prime
},m}(k,p),
\end{align}
where
\begin{align}
\tilde{V}_{ll^{\prime },m}(k,p) &=C_{3}\sqrt{\frac{2l^{\prime }+1}{2l+1}}C_{l^{\prime }020}^{l0}C_{l^{\prime }m20}^{lm}v_{3,ll^{\prime }}(k,p) +C_{6}\sqrt{\frac{2l^{\prime }+1}{2l+1}}(7\delta_{ll^{\prime }}-5C_{l^{\prime }020}^{l0}C_{l^{\prime}m20}^{lm}-2C_{l^{\prime }040}^{l0}C_{l^{\prime }m40}^{lm})v_{6,ll^{\prime }}(k,p)
\end{align}%
with
\begin{equation}
v_{s,ll^{\prime }}(k,p)=\int_{0}^{\infty }r^2dr\frac{1}{%
r^{s}}j_{l}(kr)j_{l^{\prime }}(pr).\label{vsll}
\end{equation}
For $s=3$ and odd $l$ and $l^{\prime }$, we have
\begin{align}
v_{3,ll^{\prime }}(k,p)=\left\{
\begin{array}{ll}
\frac{\pi \Gamma \left(\frac{l+l^{\prime }}{2}\right){}_{2}F_{1}\left(\frac{l^{\prime }-l-1}{2}
,\frac{l+l^{\prime }}{2},\frac{3}{2}+l^{\prime },\frac{p^2}{k^2}\right)}{8\Gamma
\left(\frac{3+l-l^{\prime }}{2}\right)\Gamma \left(\frac{3}{2}+l^{\prime }\right)}\left(\frac{p}{k}\right)^{l'},&\mbox{ for }k>p, \\
\frac{\pi \Gamma \left(\frac{l+l^{\prime }}{2}\right){}_{2}F_{1}\left(\frac{l-l^{\prime }-1}{2}%
,\frac{l+l^{\prime }}{2},\frac{3}{2}+l,\frac{k^2}{p^2}\right)}{8\Gamma \left(\frac{3+l^{\prime }-l}{2}\right)\Gamma \left(\frac{3}{2}+l\right)}\left(\frac{k}{p}\right)^{l},&\mbox{ for }p>k,
\end{array}%
\right.
\end{align}%
which is regular. However, for $s=6$ and odd $l$ and $l'$, $v_{6,ll^{\prime }}(k,p)$ becomes divergent when $l=l'=1$. To properly treat this divergence, we introduce a short-range cutoff $r_{\mathrm{UV}}$ on the lower integration limit in Eq.~\eqref{vsll}, which leads to
\begin{align}
v_{6,ll^{\prime }}(k,p) =\frac{kp}{9r_{\mathrm{UV}}}\delta _{l1}\delta _{l^{\prime }1}+\left\{
\begin{array}{ll}
\frac{\pi \Gamma \left(\frac{l+l^{\prime }-3}{2}\right){}_{2}F_{1}\left(\frac{l^{\prime }-l}{2}%
-2,\frac{l+l^{\prime }-3}{2},\frac{3}{2}+l^{\prime },\frac{p^2}{k^2}\right)}{%
64\Gamma \left(3+\frac{l-l^{\prime }}{2}\right)\Gamma \left(\frac{3}{2}+l^{\prime }\right)}k^{3}\left(%
\frac{p}{k}\right)^{l^{\prime }},&\mbox{ for }k>p, \\
\frac{\pi \Gamma \left(\frac{l+l^{\prime }-3}{2}\right){}_{2}F_{1}\left(\frac{l-l^{\prime }}{2}%
-2,\frac{l+l^{\prime }-3}{2},\frac{3}{2}+l,\frac{k^2}{p^2}\right)}{64\Gamma \left(3+%
\frac{l^{\prime}-l}{2}\right)\Gamma \left(\frac{3}{2}+l\right)}p^{3}\left(\frac{k}{p}\right)^{l},&\mbox{ for }p>k.
\end{array}%
\right. \label{v6ll}
\end{align}
We use Eq.~\eqref{v6ll} throughout our calculations and take the limit $r_{\mathrm{UV}}\rightarrow 0$ as the final step to obtain the result.

\end{document}